\documentclass{PoS}

\title{Improvement of the pion spectrum with HYP-smeared staggered fermions}

\ShortTitle{Update on pion spectrum}

\author{\speaker{Taegil Bae}, David Adams, Hyung-Jin Kim,
Jongjeong Kim, Kwangwoo Kim, and Weonjong Lee \\
Frontier Physics Research Division and
Center for Theoretical Physics, \\
Department of Physics and Astronomy, 
Seoul National University, Seoul, 151-747, South Korea \\ 
E-mail: \email{wlee@phya.snu.ac.kr}}

\author{Chulwoo Jung \\ 
Physics Department, Brookhaven National Laboratory, 
Upton, NY11973, USA \\
E-mail: \email{chulwoo@bnl.gov}}

\author{Stephen R.~Sharpe \\ 
Department of Physics, University of Washington, Seattle,
WA 98195-1560, USA \\ 
E-mail: \email{sharpe@phys.washington.edu}}

\abstract{ We extend our previous study of taste-symmetry breaking
using HYP-smeared staggered fermions in two ways. 
First, we improve the statistics
of a comparison of unimproved and HYP-smeared staggered fermions
on quenched lattices with $a\approx 0.1\;$fm. This allows us to
obtain a signal for all pion tastes, rather than just a subset,
and thus to make a complete comparison.
In addition, it allows us to differentiate
between wall and local sources.
Second, we compare HYP-smeared valence quarks to asqtad
  valence quarks on 2+1 flavor unquenched MILC lattices.
We find that taste breaking is substantially reduced by HYP-smearing,
bringing the size of this discretization effect (which is the
dominant such effect with staggered fermions) down to the size
expected generically for any fermion type.}

\FullConference{The XXV International Symposium on Lattice Field Theory\\ 
                July 30 - August 4 2007\\
                Regensburg, Germany}

\begin{document}
\section{Introduction}
\label{sec:intro}
%
%
%
Unimproved staggered fermions suffer from very large discretization errors,
particularly those that involve the breaking of taste symmetry.
Reducing these errors is crucial for practical applications,
and one popular choice is the asqtad action~\cite{ref:asqtad:1}
which has been extensively used by the MILC collaboration.
This is a tree-level $O(a^2)$ improved action, and has been
found to significantly reduce taste breaking.
Nevertheless, taste-breaking remains the dominant discretization
error, and further improvement is desirable.
This can be accomplished by the use of multiple levels
of link-smearing (with reunitarization at each stage)~\cite{ref:hyp:1,ref:HISQ}.
A tree-level $O(a^2)$ improved action with multi-level smearing
is the HISQ action~\cite{ref:HISQ}---and this is the present staggered
action of choice for charm quarks.
For light quarks, however, where $a^2 m_q^2$ effects are likely small,
we are pursuing a simpler option. This is the ``HYP action'', in
which one uses HYP-smeared links (which have three levels
of smearing)~\cite{ref:hyp:1}, but otherwise keeps the unimproved
staggered action. This action is not fully $O(a^2)$ improved, but 
our hope is that it is effective at reducing the
dominant taste-breaking discretization error 
down to the generic size, namely $a^2 \Lambda_{\rm QCD}^2$.

Last year we presented our first results from a comparative study
of taste-breaking with unimproved, HYP-smeared and asqtad valence 
fermions~\cite{ref:wlee:2}. We used the splitting in the pion multiplet
as a non-perturbative measure of taste breaking.
We found that for quenched configurations with $a\approx 0.1\;$fm
there was a dramatic reduction in taste-breaking. Indeed, our
results indicated that, rather than treating the splittings
as a leading order effect in a joint chiral-continuum expansion,
i.e. $a^2\sim p^2$, they should be treated as a next-to-leading-order
effect, i.e. $a^2 \sim p^4$. This would substantially simplify
the formulae of staggered chiral perturbation 
theory~\cite{ref:wlee:1,ref:bernard:1}.

During the last year we have improved the statistics of the quenched calculation,
allowing a more thorough calculation of the pion spectrum, now including
all tastes, although the overall conclusions are unchanged.
More importantly, we have calculated the pion spectrum
with the valence HYP action on unquenched
MILC ``coarse'' lattices ($a\approx 0.125\;$fm), allowing a direct
comparison with results using valence asqtad action, and also allowing
us to determine the power-counting appropriate for the HYP action.

We do not repeat here the technical discussion 
concerning the types of sources and sinks that we use, 
refering the interested reader to last year's talk~\cite{ref:wlee:2}.
We focus instead on the new numerical results and 
the corresponding conclusions.
Further details will be given in an
upcoming publication.

\section{Update on quenched results}
\label{sec:quenchedupdate}

We use quenched lattices of size $16^3\times 64$
generated with the Wilson gauge action at coupling
$\beta=6$, so that $1/a\approx 1.95\;$GeV.
Compared to last year,
we have increased our statistics from 218 to 370 configurations.
For unimproved fermions we use quark masses
$a m_q =0.005$, $0.01$, $0.015$, $0.02$, $0.025$ and $0.03$,
while for HYP fermions we use
$a m_q = 0.01$, $0.02$, $0.03$, $0.04$, and $0.05$.
Despite appearances, these ranges corresponds to heavier
physical masses for unimproved than for HYP-smeared fermions.
This is because $Z_m \approx 2.5$ for the former, while
$Z_m\approx 1$ for the latter.
For example, the physical strange quark mass 
is $a m_s^{\rm phys} \approx 0.025$ for unimproved fermions
and $0.052$ with HYP smearing.

The increase in statistics allows us to calculate
the masses of pions of all tastes. Last year we had
results only for the ``LT'' (local in time) tastes, i.e.
$\xi_5$, $\xi_4$, $\xi_{i5}$ and $\xi_{i4}$.
For pseudoscalar spin, $\gamma_S=\gamma_5$,
these are the tastes created
by an operator that is local in time.
This year we have signals also for the ``NLT'' (non-local in time) tastes
${\bf 1}$, $\xi_i$, $\xi_{45}$ and $\xi_{ij}$.
These are produced by our sources (which reside on a single
timeslice) by their coupling to the axial current,
$\gamma_S=\gamma_4\gamma_5$, which is suppressed 
relative to the production of LT tastes.

As described last year, we use two choices of sink operators,
and find them to be indistinguishable. All results presented
here are obtained with the ``Golterman''~\cite{ref:golterman:1} sinks.
We also use two sources---an extended ``cubic wall source'' and
a local ``cubic U(1) source''. We find that both work comparably
well for LT pions, while 
the cubic wall source
leads to smaller errors for the NLT pions, although the
improvement is smaller for the HYP action than for the unimproved action.

\begin{figure}[t!]
\includegraphics[width=0.5\textwidth]{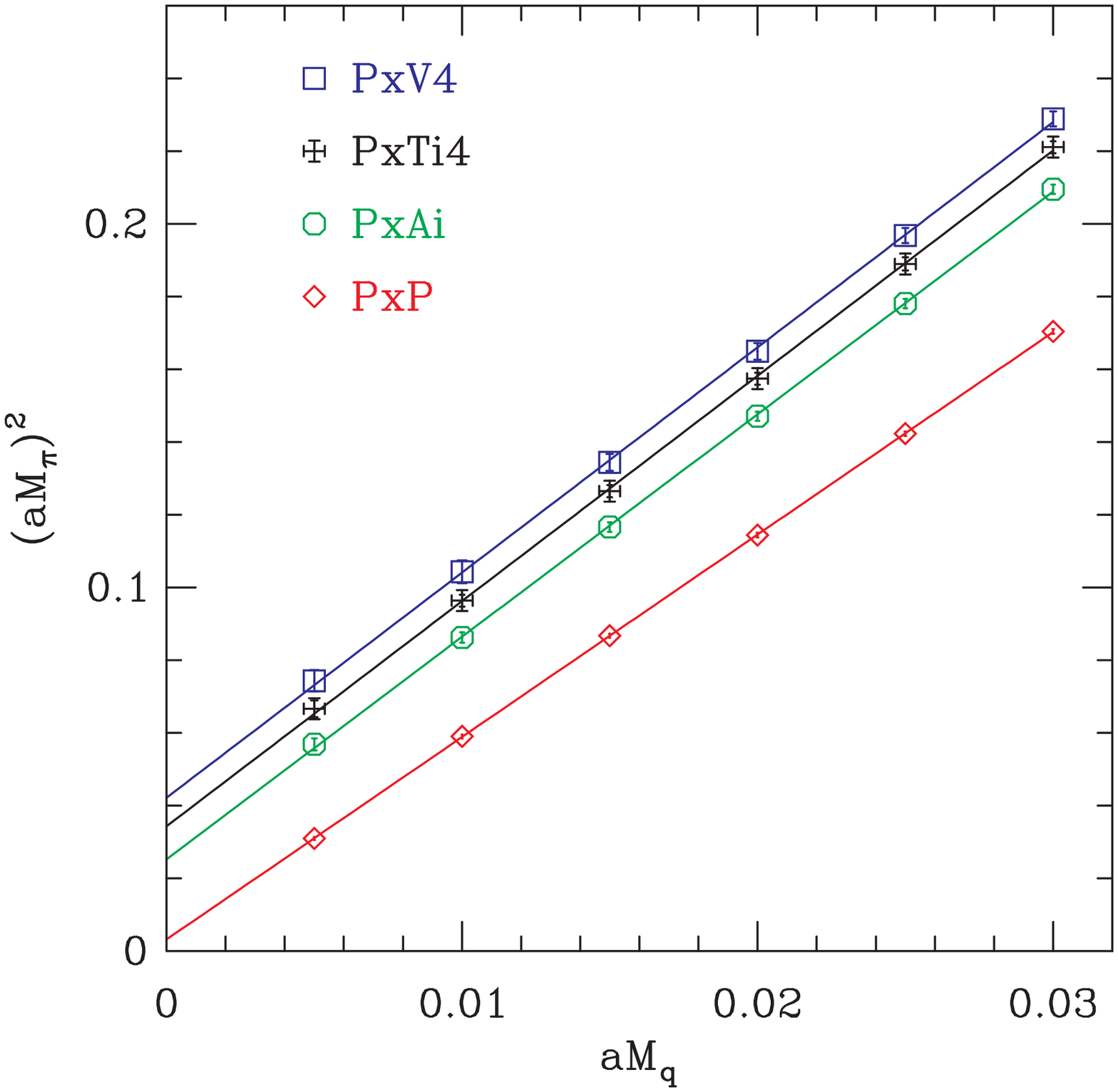}
\includegraphics[width=0.5\textwidth]{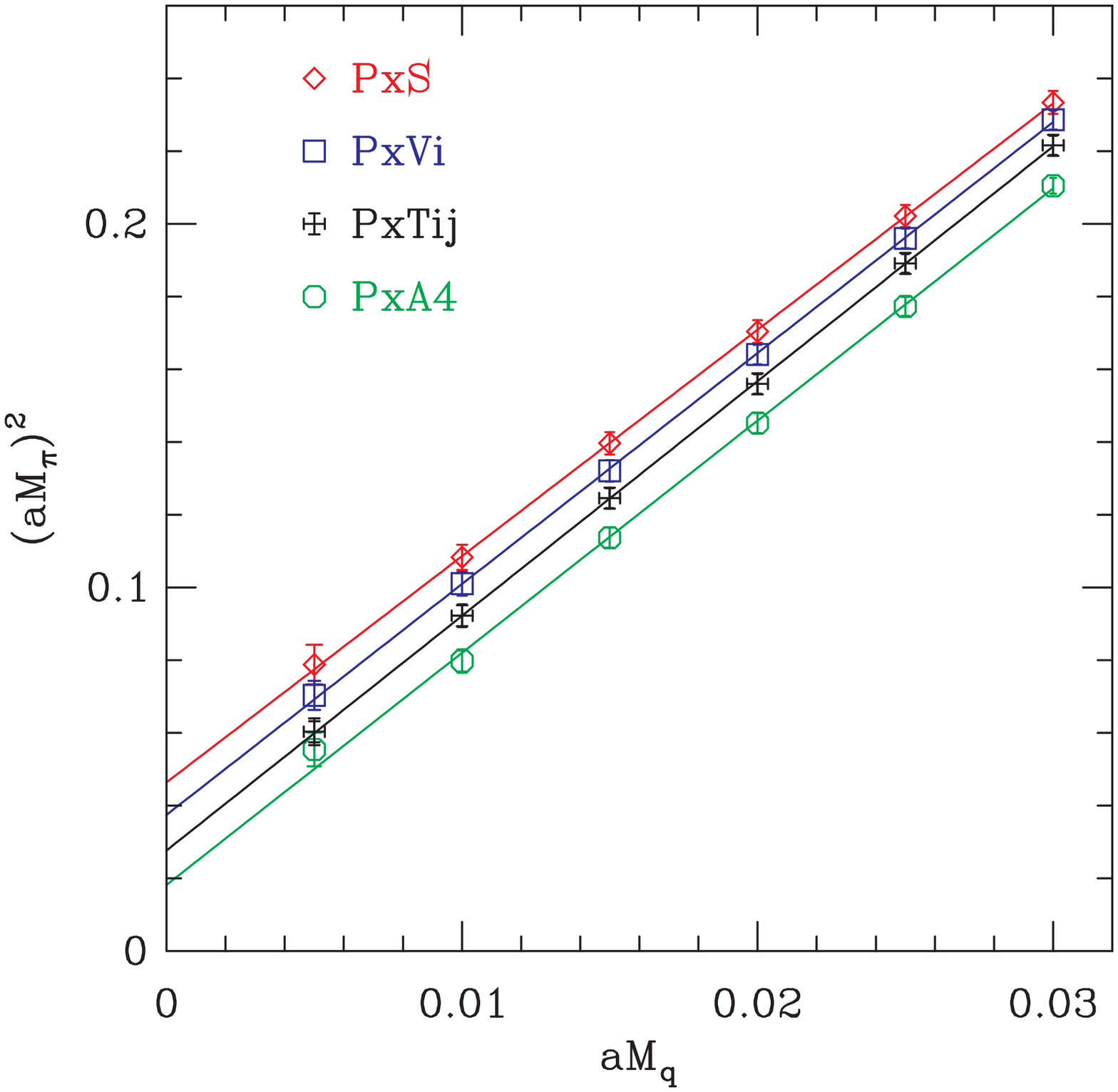}
\caption{$(a m_\pi)^2$ vs.~$am_q$ for unimproved staggered fermions
  with cubic wall sources on quenched lattices.
  Left panel: LT tastes; Right panel: NLT tastes.
  Linear chiral extrapolations are shown.}
\label{fig:unimp-gol}
\end{figure}

\begin{figure}[t!]
\includegraphics[width=0.5\textwidth]{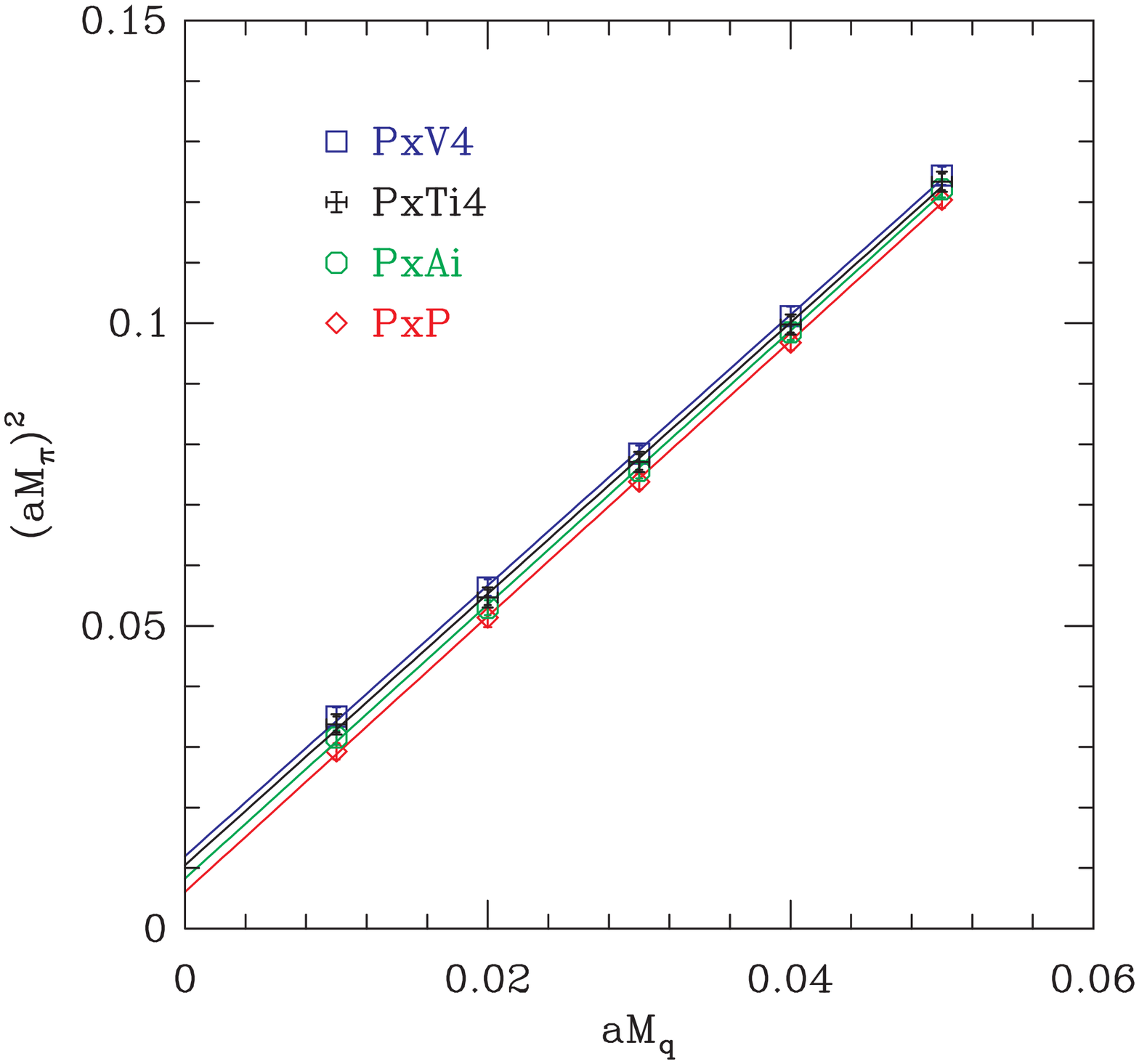}
\includegraphics[width=0.5\textwidth]{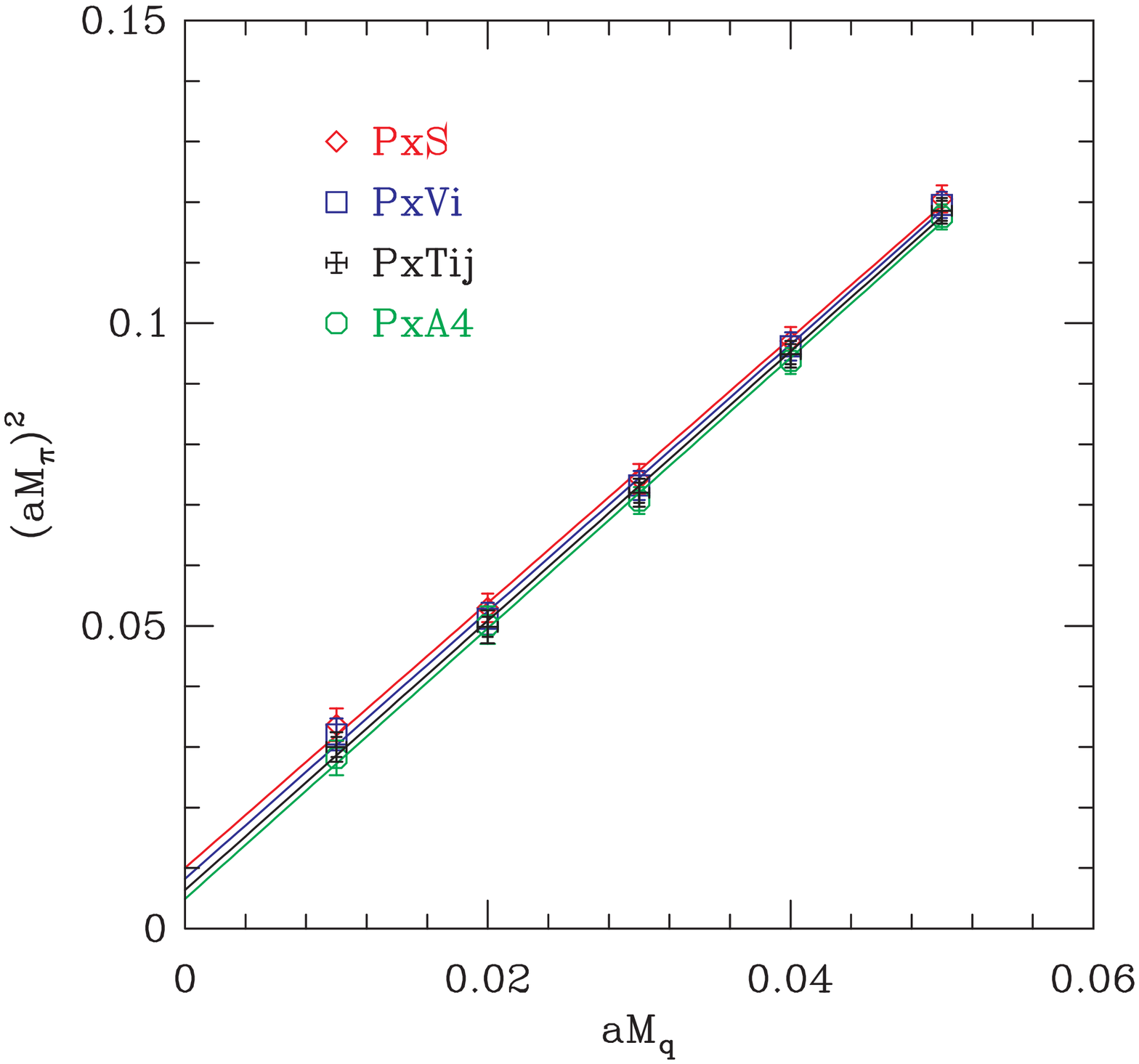}
\caption{$(am_\pi)^2$ vs.~$a m_q$ for   HYP-smeared
   staggered fermions with cubic wall sources on quenched lattices:
  Left panel: LT tastes; Right panel: NLT tastes. 
Linear chiral extrapolations are shown.}
\label{fig:hyp-gol}
\end{figure}

In Figs.~\ref{fig:unimp-gol} and \ref{fig:hyp-gol} we show, respectively, 
the pion spectrum for unimproved and HYP-smeared staggered fermions.
The dramatic reduction in the breaking of taste symmetry observed
last year is seen to carry over to the NLT tastes. For all except
the taste singlet, this result is expected
because of the predicted approximate SO(4) symmetry, which holds
at leading order in staggered chiral perturbation theory~\cite{ref:wlee:1}.
The SO(4) symmetry combines tastes $\xi_i$ (NLT) and $\xi_4$ (LT)
into a single multiplet with taste $\xi_\mu$, and similarly
for the SO(4) irreps with tastes $\xi_{5\mu}$ and $\xi_{\mu\nu}$.
Of the NLT tastes, only the taste singlet is unpaired, and thus 
the fact that its mass is very close to those of the other tastes
does not follow from SO(4) symmetry, but rather is an additional,
independent indication of the smallness of taste breaking with HYP
fermions.

In Table~\ref{tab:delta-hyp-unimp}, we list the
mass-squared splittings after linear
extrapolation to the chiral limit. These are proportional
to $a^2$ in staggered chiral perturbation theory. Note that
knowledge of $Z_m$ is not needed in order to compare these
quantities between the two types of fermion.
Our main conclusion is that taste breaking is reduced by
an order of magnitude by HYP smearing. 
In fact, only for the LT tastes (the left columns) are the
splittings significantly different from zero with HYP 
smearing. For the NLT tastes, even the ordering 
of states is not clear.
We also note that there is evidence for SO(4) symmetry breaking
for unimproved fermions at the $2-3\sigma$ level, 
with the NLT tastes systematically lower than their LT partners.
We are investigating this issue further.

%
\begin{table}[h!]
\begin{center}
\begin{tabular}{c c c| c c c}
\hline
Taste ($F$) & $\Delta_F$ (unimproved) & $\Delta_F$ (HYP) &
Taste ($F$) & $\Delta_F$ (unimproved) & $\Delta_F$ (HYP) 
\\
\hline
$\xi_{i 5}$ & 0.0230(09) &  0.0022(09)  &
$\xi_{4 5}$ & 0.0200(34) &  -0.0012(29)
\\
$\xi_{i 4}$ & 0.0360(24) &  0.0044(10)  &
$\xi_{i j}$ & 0.0245(40) &  0.0003(31) 
\\
$\xi_{4}  $ & 0.0447(27) &  0.0059(22) &
$\xi_{i}  $ & 0.0342(42) &  0.0022(30) 
\\
& & & $I        $ & 0.0432(44) &  0.0040(30) \\
\hline
\end{tabular}
\end{center}
\caption{Comparison of taste symmetry breaking with unimproved
and HYP-smeared fermions on quenched lattices with cubic wall sources.
Here $\Delta_F = [a m_\pi(F)]^2 - [a m_\pi(\xi_5)]^2$,
extrapolated to the chiral limit.}
\label{tab:delta-hyp-unimp}
\end{table}

In conclusion, we confirm and extend the findings of last year.
Unimproved staggered fermions show taste breaking that
is of the same order as the light pion masses, so that
${\cal O}(a^2)\sim {\cal O}(p^2)$ is the appropriate power
counting. Furthermore, the slopes differ significantly
for the different tastes, indicating an important
next-to-leading-order (NLO) contribution from ${\cal O}(a^2 p^2)$ terms.
By contrast, for HYP-smeared fermions both the splittings
and the differences in the slopes are much smaller.
Thus we conclude that one should treat ${\cal O}(a^2)$ effects
as being of NLO for HYP fermions on these lattices.
The important practical question is whether this holds
also for unquenched lattices.

\section{Comparison of asqtad and HYP-smeared staggered fermions}
\label{sec:num:cmp}
To address this we have compared asqtad and HYP-smeared staggered
valence fermions on one set of the coarse 
$2+1$ flavor MILC lattices~\cite{ref:milc:2}.
The parameters of the study are given in
Table~\ref{tab:par:unquenched}.
The range of quark masses for asqtad quarks is
similar to that used above, the physical strange quark mass
$a m_s^{\rm phys}\approx 0.045$ lying near the upper end.
It turns out, however, that the HYP quarks we use are
somewhat lighter. For the same bare quark mass, the physical
quark mass is about 1.4 times smaller with HYP than with asqtad quarks.
Part of this difference is due to the normalization convention
for tadpole improvement used by Ref.~\cite{ref:milc:1}, 
which has the effect that the bare quark mass for asqtad
quarks should be multiplied by $1/u_0$ to match
the convention used for unimproved and HYP quarks.
Here $u_0$ is the ``mean link'', 
which is $\approx 0.87$ for these lattices.
The remainder of the difference 
is presumably due to a difference in the values of 
$Z_m$ for the two lattices.

%
\begin{table}[b]
\begin{tabular}{ c  c }
\hline
parameter & value \\
\hline
gauge action  & 1-loop tadpole-improved Symanzik \\
sea quarks &  $2+1$ flavors of asqtad staggered 
($am_\ell=0.01$, $am_s=0.05$) \\
$\beta$  & 6.76 \\
$a$ & 0.125 fm \\
geometry & $20^3 \times 64$  \\
\# of confs & 640 (asqtad)/ 406 (HYP) \\
valence quark type & asqtad and HYP staggered \\
valence quark mass (asqtad)& 
0.007, 0.01, 0.02, 0.03, 0.04, 0.05  \\
valence quark mass (HYP)& 
0.05, 0.01, 0.015, 0.02, 0.025, 0.03, 0.035, 0.04, 0.045, 0.05  \\
\hline
\end{tabular}
\caption{Simulation parameters used for the comparison of asqtad
and HYP valence quarks. Asqtad results are
from Ref.~\protect\cite{ref:milc:1}.}
\label{tab:par:unquenched}
\end{table}
\begin{figure}[t!]
\begin{center}
\includegraphics[width=0.5\textwidth]{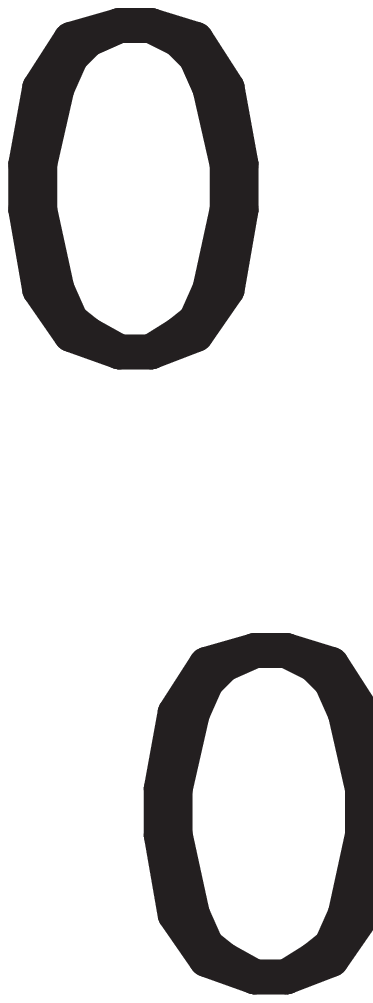}
\end{center}
\caption{$(r_1 m_\pi)^2$ vs.~$2 r_1 m_q$ for asqtad valence quarks
on unquenched lattices. From   Ref.~\cite{ref:milc:1}.}
\label{fig:mpisq-milc}
\end{figure}

In Fig.~\ref{fig:mpisq-milc}, we show the results 
obtained in Ref.~\cite{ref:milc:1} using asqtad valence quarks.
One sees that taste-breaking is significant, with
${\cal O}(a^2)\sim {\cal O}(p^2)$ being the appropriate power
counting. Note that the taste-breaking with unimproved
staggered quarks would be larger still, since these 
lattices are coarser than the quenched lattices used earlier.
Despite the relatively large taste breaking, 
discretization effects of NLO (i.e. ${\cal O}(a^2 p^2)$) are 
very small. This is shown both by the lack of 
breaking of SO(4) symmetry, and the result that
the slopes for different tastes are close.
The latter result differs from that found above with unimproved
fermions.

\begin{figure}[t!]
\includegraphics[width=0.5\textwidth]{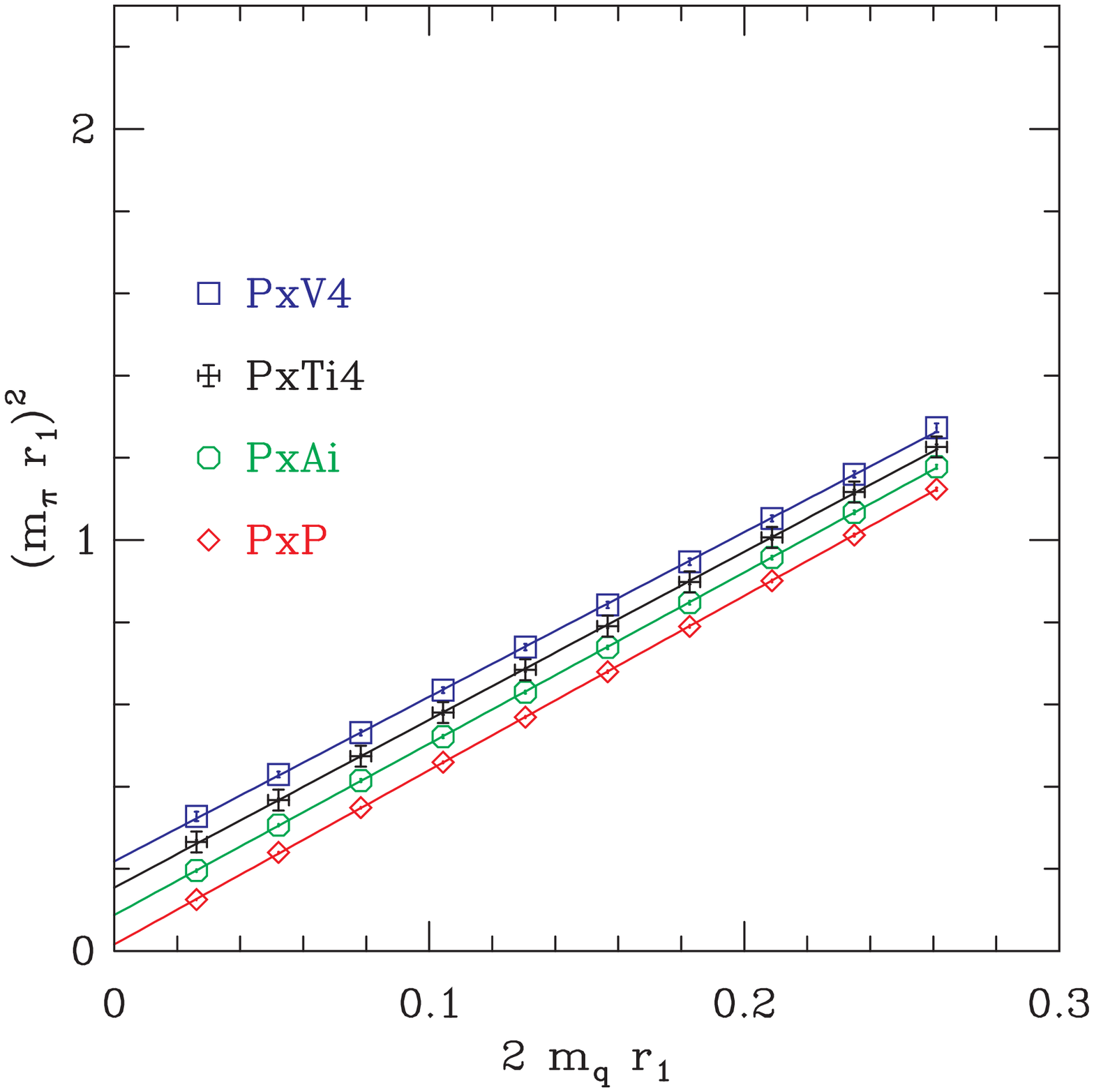}
\includegraphics[width=0.5\textwidth]{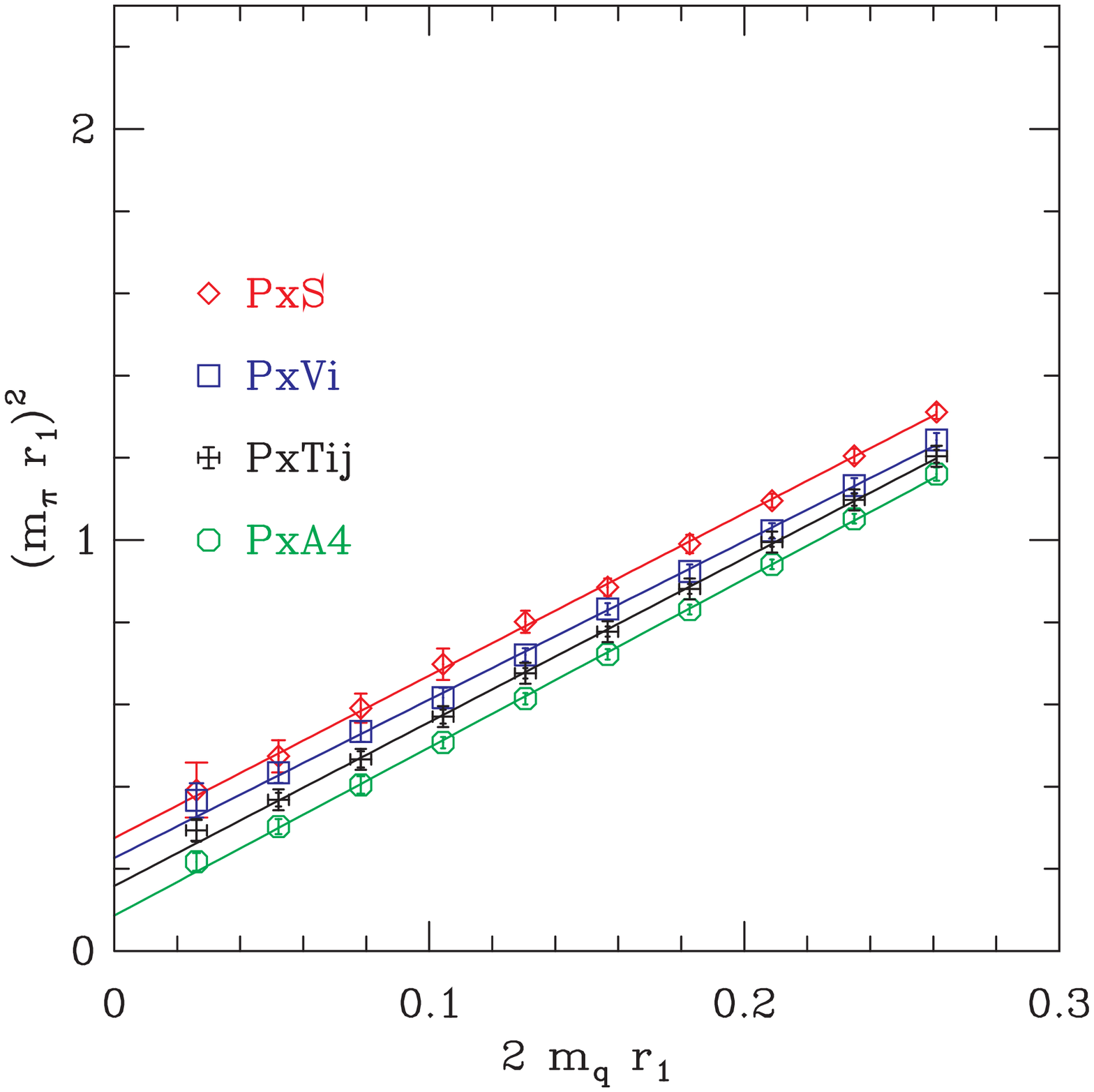}
\caption{$(r_1 m_\pi)^2$ vs.~$2 r_1 m_q$ for HYP valence quarks
on unquenched lattices, using cubic U(1) sources.
Left panel: LT tastes; Right panel: NLT tastes.}
\label{fig:mpisq-mix-gol-NLT}
\end{figure}
Our new results using HYP-smeared staggered quarks are
shown in Fig.~\ref{fig:mpisq-mix-gol-NLT}.
Results are with cubic U(1) sources; those with cubic
wall sources are not yet available.
We plot using the same dimensionless variables
as in Fig.~\ref{fig:mpisq-milc}. The difference
in slopes between asqtad and HYP fermions 
(by a factor of about 1.4) is due to the difference in quark
mass conventions and renormalization factors discussed above.
Irrespective of this detail, we see that taste breaking
is substantially reduced for HYP-smeared quarks.
To make this quantitative we quote,
in Table~\ref{tab:delta:unquenched},
the splittings in the chiral limit (using $r_1$ to
set the scale, rather than the $1/a$ used above).
We see that HYP smearing reduces taste breaking
by a factor of 2.5-3, although the
 ordering of states is the same
for both asqtad and HYP actions.
The improvement in taste-breaking is
similar to that found using the HISQ action~\cite{ref:HISQ}.

\begin{table}[h!]
\begin{center}
\begin{tabular}{ccc|ccc}
\hline
Taste ($F$) & $\Delta_F$ (asqtad) & $\Delta_F$ (HYP) &
Taste ($F$) & $\Delta_F$ (asqtad) & $\Delta_F$ (HYP) \\
\hline
$\xi_{i 5}$    & 0.205(2) &  0.072(2)  &
$\xi_{45}$     & 0.205(2) &  0.070(18) \\
$\xi_{i4}$     & 0.327(4) &  0.138(4) &
$\xi_{ij}$     & 0.327(4) &  0.142(20) \\
$\xi_{4} $     & 0.439(5) &  0.203(7) &
$\xi_{i}$      & 0.439(5) &  0.210(28) \\
& & & $I         $     & 0.537(15) & 0.259(39) \\
\hline
\end{tabular}
\end{center}
\caption{Taste symmetry breaking effect:
$\Delta_F = [r_1 m_\pi(F)]^2 - [r_1 m_\pi(\xi_5)]^2$
extrapolated to the chiral limit.
Results from unquenched lattices comparing valence
asqtad (from Ref.~\cite{ref:milc:1}---with SO(4) 
symmetry enforced) to HYP-smeared
fermions (with cubic U(1) sources).}
\label{tab:delta:unquenched}
\end{table}

We can characterize the taste-breaking errors in terms of
a non-perturbative scale. Using the 
splitting of the ``2-link'' pions (taste $\xi_{\mu\nu}$) from
the Goldstone pion as an ``average'' splitting, and setting
$\Delta a^2 m_\pi^2 = a^4 \Lambda^4$, we find $\Lambda\approx0.6\;$GeV.
This should be compared to scales $\approx 1\;$GeV for unimproved
staggered fermions. Thus improvement has reduced discretization
errors to the expected generic size, i.e. $\Lambda\sim \Lambda_{\rm QCD}$.

Looking in more detail at the results for HYP fermions, we note
that SO(4) breaking is small, and that the slopes of
$m_\pi^2$ versus $m$ graphs are similar for all tastes.
Thus, as for the asqtad quarks, the NLO 
${\cal O}(a^2 p^2)$ terms responsible for these effects
must be small. 

Despite the reduction in taste breaking, 
it is clear from the figures that, for this range
of quark masses (which is the range we are using for
calculating matrix elements), the appropriate power counting 
for the MILC coarse lattices remains ${\cal O}(a^2)\sim {\cal O}(p^2)$
for HYP fermions.

\section{Conclusions}

The most important conclusion for future work is that HYP smearing
leads to a significant reduction in taste breaking compared to the
asqtad action. This was expected from previous studies, but it is
important to demonstrate this on the unquenched MILC lattices that we
are using for calculating weak matrix elements.  The reduction is,
however, not small enough to change the power counting for staggered
chiral perturbation theory on the coarse lattices. If, however, one
extrapolates to the fine MILC lattices, on which one expects that
taste breaking will be reduced by about 2.5 (which is the relative
size of $\alpha^2 a^2$), then it seems likely that one will be able to
treat the discretization errors as a NLO effect, i.e. that ${\cal
O}(a^2)\sim {\cal O}(p^4)$.  Pictorially, the spectrum on the fine
lattices should look like that for the quenched lattice results of
Fig.~\ref{fig:hyp-gol}.  This would lead to a significant
simplification in the fitting forms from staggered chiral perturbation
theory.

\section{Acknowledgment}
\label{sec:ack}
W.~Lee acknowledges with gratitude that the research at Seoul National
University is supported by the KICOS grant K20711000014-07A0100-01410,
by the KRF grant KRF-2006-312-C00497, by the BK21 program, and by the
US DOE SciDAC-2 program.
The work of S.~Sharpe is supported in part by the US DOE grant
no. DE-FG02-96ER40956, and by the US DOE SciDAC-2 program.


%
%
%

\end{document}